\documentclass[aps,prb,twocolumn]{revtex4}
\usepackage{epsfig}
\usepackage{color}
\usepackage{amsmath}
\usepackage{amssymb}
\usepackage{bbm}

\newcommand{\be}{\begin{equation}}
\newcommand{\ee}{\end{equation}}
\newcommand{\bea}{\begin{eqnarray}}
\newcommand{\eea}{\end{eqnarray}}

\begin{document}

\title{Sequence of superconducting states in field cooled $FeCr_2S_4$ }

\author{Naoum Karchev}

\affiliation{Department of Physics, University of Sofia, 1164 Sofia, Bulgaria}

\begin{abstract}
In the present article we discuss theoretically the emergence of superconductivity in field cooled $FeCr_2S_4$. The chromium electrons form a triplet $t_{2g}$ states and due to antiferromagnetic exchange with the iron spins have Zeeman splitting. Applied, during preparation, magnetic field along the moment of iron ions, successively compensates the Zeeman splittings. The chromium electrons with zero Zeeman energy form Cooper pairs induced by iron magnons. In that way, we predict theoretically the existence of sequence of superconducting states in field cooled $FeCr_2S_4$. Actually there are three different superconductors prepared applying, during preparation, different magnetic fields. In these compounds superconductivity coexist with the saturated magnetism of iron ions.	

\end{abstract}

%\pacs{75.50.Gg,74.20.Mn,74.20.Rp}

\maketitle

The synthesis of novel superconductors has been attracting attention for a long time.
One way to do this is to subject a suitable material to strong hydrostatic pressure. The coexistence of  ferromagnetism and superconductivity in $UGe_2$ under pressure was reported 
twenty years ago, in the paper \cite{UGe2000}. The invention triggered a very intense experimental and theoretical study of the phenomenon \cite{Huxley2001,Tateiwa2001b,Motoyama2001,Huxley2002,Pfleiderer2009,Aoki2019}. 

In his theoretical studies Ashcroft predicted that metalized hydrogen \cite{Ashcroft68} or hydrogen rich alloys \cite{Ashcroft04} can possess high temperature superconductivity. Ashcroft's idea was supported in numerous experiments. An important discovery leading to room-temperature
superconductivity is the pressure-driven hydrogen sulfide with a confirmed transition temperature of 203 K at 155 GPa \cite{Drozdov15}. The most recent examples of a metal hydride are lanthanum hydride which has $Tc = 250 - 260 K$ at $ 180 - 200 GPa $ \cite{Drozdov19,Somayazulu19} and sulfur hydride with room-temperature $T_c =287 K$ achieved at $287GPa$ \cite{Snider20} .

Another way to fabricate unconventional superconductor is by chemical manipulation. The most famous example is copper-oxide superconductor.  The parent compound $La_2CuO_4$ is  an antiferromagnetic Mott insulator with N\'{e}el  temperature  $T_N = 300 K$.  The parent compound can be doped by substituting some of the trivalent $La$ by divalent $Sr$. The result is that
$x$ holes are added to the $Cu-O$ plane in $La_{2-x}Sr_xCuO_4$, which is called hole doping. The hole-doping suppresses  the antiferromagnetic order and at $x=0.03-0.05$ hole concentration
the system undergoes quantum antiferromagnetic-paramagnetic transition. After suppression of the antiferromagnet, superconductivity appears, ranging from $x=0.06-0.25$ \cite{Bednorz86}. 
The electron doping is realized in the compound  $Nd_{2-x}Ce_xCuO_4$ \cite{Tokura89} when $x$ electrons are added. Details are given in many review articles and books, for example \cite{Ginsberg89,Dagotto94,Anderson97,Lee06}.

The magnetic field induced superconductivity (FISC) is one more issue of special interest. Experimentally, (FISC) in the $H_{c2}-T$ phase diagram was observed in $Eu_xSn_{1-x}Mo_6S_8$ \cite{Fischer75}. The domain of superconductivity in $Eu_0.75Sn_0.25Mo_6S_8$ extends from 4 to 22 T at $T=0$ and from $T=0$ to $T=1 K$ at $H = 12 T$ \cite{Fischer84}. The magnetic-field induced superconductivity was attributed to Jaccarino-Peter (JP) compensation mechanism\cite{JP62}. The idea is that in a ferromagnetic metal the conduction electrons are in an effective field due to the exchange interaction with the localized spins. It is in general so large as to inhibit the occurrence of superconductivity. For some systems the exchange interaction have a negative sign. This  allows for the conduction electron polarization to be canceled by an external magnetic field so that if, in addition these metals possess phonon-induced attractive electron-electron interaction, superconductivity  occurs in the compensation region. In more complicated cases superconductivity can occur in two domains: one extends from zero applied magnetic field to small pair-breaking field which suppress superconductivity, and the other at the high field in the compensation region \cite{Fischer84}.

A great deal of interest has been centered on the heavy fermions in cerium and uranium systems. The heavy-fermion system $CePb_3$ at zero field is an antiferromagnet. In \cite{Lin85} the authors report  magnetic field of 14 T induces the system into the superconducting state below 0.20 K. Similarly, at 0.48 K, 15 T magnetic field  drives the sample superconducting. 
The (FISC) in these compounds is considered to be due to the Jaccarino-Peter mechanism, extended to antiferromagnetic superconductors \cite{Shimahara02}. 

$URhGe$ displays ferromagnetism with magnetic moment oriented along the c -axis, and spin-triplet superconductivity at a lower temperature \cite{Hardy05}.
In an external magnetic field along the b-axis perpendicular to c-axis, superconductivity disappears at about $H=2 T$. However, at higher magnetic fields,
in the range from $8 T$ to $13.5 T$, it reappears again \cite{Sheikin05}.

Finally, magnetic-field-induced superconductivity has been observed in organic superconductors  \cite{Uji01,Balicas01,Konoike04}.

It is important to emphasize, that magnetic-field induced superconductivity disappears when the field is switched off.

In the present article we investigate two sub-lattice $FeCr_2S_4$ spinel. The sub-lattice A sites are occupied by $F^{2+}$ (s=2) iron ions, while sub-lattice B sites are occupied by $Cr^{3+}$ (s=3/2) chromium ions. The $Fe^{2+}$ and $Cr^{3+}$ ions are located at the center of tetrahedral and octahedral  $S^{2-}$ cages, respectively, and the three $Cr^{3+}$ electrons occupy the lower energy $t_{2g}$ bands. 

The shape of magnetization-temperature diagrams for all spinels is remarkable and emblematic. It shows that the system has two phases. At low temperature $(0,T^{*})$, where $T^*$ is the temperature at which the magnetization is maximal, the magnetic orders of the $A$ and $B$ spins  contribute to the magnetization of the system, while at the high temperature  $(T^{*},T_N)$, only the sub-lattice A spins have non-zero spontaneous magnetization. At $T^{*}$ the system undergoes partial-order transition. There is no additional spontaneous symmetry breaking and the Goldstone boson has a ferromagnetic dispersion in both phases.

Partial order is well known phenomenon and has been subject to extensive studies. There are exact results for the partially ordered systems \cite{Diep87,Larkin66,Diep04}. Frustrated antiferromagnetic systems has been studied by means of Green function formalism. Partial order and anomalous temperature dependence of specific heat have been predicted \cite{Diep97}. Experimentally the partial order  has been observed in $Gd_2Ti_2O_7$ \cite{POexp04}. Monte Carlo method has been utilized to study the nature of partial order in Ising model on  $kagom\acute{e}$ lattice \cite{Diep87}. Spin-wave theory of partial order is developed in \cite{Karchev08}.

\begin{figure}[!ht]
%\centering\includegraphics[width=3.7in]{(fig1)Sc-FeCrS-M-T.pdf}
\epsfxsize=\linewidth
\epsfbox{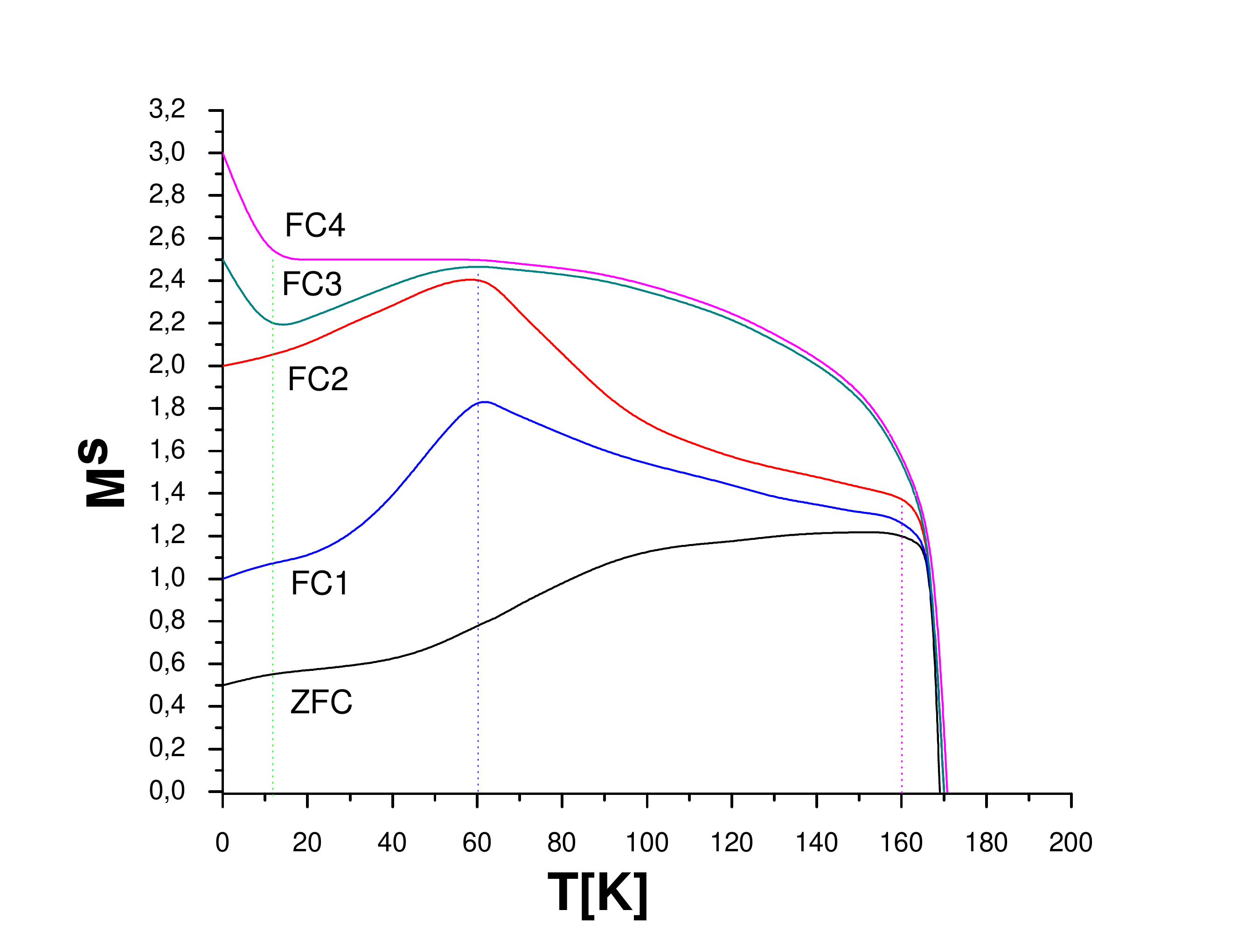}
\caption{(Color online) The temperature dependence of the spontaneous magnetization $M^s$ of ZFC, FC1, FC2, FC3 and FC4  ($FeCr_2S_4$) spinel. The red, vertical line indicates the temperature ($160 K$) at which the first chromium electron onsets the magnetization, the blue line ($60 K$) -the second electron an the green one ($10 K $) the third. }\label{(fig1)}
\end{figure}

We focus on field cooled $FeCr_2S_4$. The material is field cooled (FC) if, during the preparation, an external magnetic field is applied upon cooling. If the applied field is below $0.01 T$ it is zero field cooled (ZFC). The magnetization-temperature and magnetic susceptibility curves for (ZFC) and (FC) spinel show a remarkable difference below N\'{e}el $T_N$ temperature 
\cite{spinel+,spinelFeCr2S4,spinelCv1,spinel08,spinel++,spinel11b,spinel11a,spinel11c,spinel12a,spinel12b,spinel+1,spinel+2,spinel+3}. In the case of $FeCr_2S_4$ spinel the curves, which show the temperature dependence of spontaneous magnetization $M^s$, are depicted in Fig.\ref{(fig1)}. 

%\vskip -1.5cm

The ZFC, FC2 and FC3 curves are adapted from experimental results \cite{Yang00,Tsurkan10,Yto11,Lin14} and the rest ones, FC1 and FC4, are phenomenological  extrapolations. The N\'{e}el  temperature is $T_N = 170 K$. For ZFC spinel the temperature, at which the magnetization is maximum, is  $T^*=160 K $. The system undergoes a partial-order transition from the high temperature  $(T^{*},T_N)$ phase, where only the iron ions have non-zero spontaneous magnetization, to low temperature one $(0,T^{*})$, where both the iron and chromium ions have non-zero spontaneous magnetization. The exchange constant of chromium and iron spins is antiferromagnetic, because of which the magnetization increases below $T_N$ and decreases below $T^{*}$.  
The subtle point is that the onset of magnetism of the three  $Cr^{3+}$ ($t_{2g}$) electrons is at different temperatures.
One of them starts to form magnetic order below $T^{*}$. The ZFC curve in (Fig.\ref{(fig1)}), shows small dip at $10 K$ which indicates that another chromium electron starts to contribute the magnetic order of the system at this temperature. The contribution of the third  $Cr^{3+}$ electron becomes more clear in the experiment with FC spinel-curve FC2.
It shows that applied during the preparation magnetic field exceeds the Zeeman splitting energy of the chromium electron and its magnetic order becomes parallel to the iron magnetic moment. As a result, below $T^{*}$ the system undergoes partial-order transition, the chromium electron onset magnetic order and the spontaneous magnetization of the system increases and reaches the maximum at $60K$. It undergoes a second partial-order transition to system where the second  $Cr^{3+}$ electron onset magnetic order anti-parallel to the iron one  and the spontaneous magnetization of the system decreases. At $10 K$ the curve has neither dip nor increase, which means that applied magnetic field compensates Zeeman splitting of the third  chromium electron. The FC1 curve in the middle is an extrapolation for the case when the magnetic field applied during the preparation compensates the Zeeman splitting energy of the $ Cr ^ {3 +} $ electron. Therefore, in the temperature range $ (60 K-170 K) $ only iron electrons contribute to the magnetization of the system. Increasing the applied, during preparation, magnetic field one obtains the magnetization-temperature curve FC3. It has two characteristic features: first the onset of the magnetism of iron and the first chromium electron is at the same temperature $T_N=170 K$, second the applied magnetic field exceeds the Zeeman splitting energy of the third chromium electron and its magnetic order becomes parallel to the iron magnetic moment. As a result, below $T=10 K$ the system undergoes partial-order transition, the third chromium electron onset magnetic order and the spontaneous magnetization of the system increases. The last curve FC4 is an extrapolation for the case when the applied field compensates Zeeman splitting of the second $ Cr ^ {3 +} $ electron.

When, during the preparation of the $FeCr_2S_4$ spinel, magnetic field is applied along the iron magnetization and switched off when the process is over there are two important consequences: i) Iron electrons are localized, and the applied field do not affect iron magnetization. The experimental evidences for this is the part of the curves near $T_N=170 K$ in  Fig.\ref{(fig1)} which is not affected by the applied magnetic field. Therefore we have to study the spin fluctuations of iron electrons by means of Heisenberg model without applied magnetic field. 
 ii) The magnetic field applied along the iron magnetization decreases the Zeeman splitting of chromium electrons and their contribution to the magnetization of the system decreases. This remains true when the applied field is switched off. To describe theoretically the phenomenon a "frozen" magnetic field should be included in the equations for the fermion dispersion, which leads effectively to decreasing of Zeeman splitting. 

Based on the analysis of Fig.\ref{(fig1)}, we consider a spin-fermion model of $FeCr_2S_4$ spinel, with three bands describing $t_{2g}$ chromium electrons and 
spin $s=2$ operators for localized $F^{2+}$ electrons. The iron-chromium exchange constants are antiferromagnetic, different for the three chromium electrons. Magnetic field in the Hamiltonian models the decrease of the Zeeman splitting during preparation of the material. We study the appearance and disappearance of superconducting states as a function of the field. The Hamiltonian of the model is
\bea \label{FeCrS1}
 h  = & - & t\sum\limits_{\ll ij \gg _B }\sum\limits_{\sigma,l}{\left( {c_{i\sigma l }^ + c_{j\sigma l } + h.c.} \right)}-H \sum\limits_{i\in B,\, l} {S^{zB}_{il}} \nonumber \\
& + & \sum\limits_{  \langle  ij  \rangle, l } J_l\, {{\bf S_i^A}}\cdot {\bf S_{j l}^B}
-  J_A\sum\limits_{  \ll  ij  \gg_A  } {{\bf S_i^A}
	\cdot {\bf S_j^A}}, 
\eea 
where $S^{\nu B}_{il}=\frac 12\sum\limits_{\sigma\sigma'}c^+_{i\sigma l}\tau^{\nu}_{\sigma\sigma'}c^{\phantom +}_{i\sigma' l}$, with the Pauli
matrices $(\tau^x,\tau^y,\tau^z)$, is the spin of the $t_{2g}$ chromium 
electrons at the sub-lattice $B$ site , ${\bf S}_i^A$ is the spin operator of the localized iron electrons  at the sub-lattice $A$ site. The
sums are over all sites of a body centered cubic lattice, $\langle i,j\rangle$ denotes the sum over the nearest neighbors, while $ \ll  ij  \gg_A$ is a sum over all sites of sub-lattice $A$. The Heisenberg term $(J_A > 0)$ describes ferromagnetic
exchange between iron spins, and $J_l>0$ are the antiferromagnetic exchange constants between iron and chromium spins. $H>0$ is the "frozen" applied magnetic field in units of energy.

The compensation mechanism of the field induced superconductivity suggests that the formation of Cooper pairs is possible when Zeeman splitting of electrons is compensated by the applied magnetic field.
We choose the exchange constants well separated
$J_1<J_2<J_3$,
so that if the magnetic field $H$ compensates the Zeeman splitting of one of $t_{2g}$ chromium electrons, it is far from the compensation of Zeeman energy  of the other two electrons.  

With this in mind, we can simplify our study. When the value of the "frozen"  magnetic field is close to the Zeeman  energy of one of $t_{2g}$ chromium electrons, we can consider one band spin-fermion model of this electron instead of model (\ref{FeCrS1}). The contribution of dropped fermions can be accounted for by appropriate choice of the parameters.  In this way we 
consider three independent, one band spin-fermion models. In momentum space representation, the Hamiltonians $h_l$ ($l=1,2,3$) have the form   
\bea\label{FeCrS2} \nonumber
h_l & = &  \sum\limits_{k\in B_r} \varepsilon_k a_k^+ a_k  + \sum\limits_{k\in B_r \sigma} \varepsilon_{k \sigma l} c_{k \sigma l}^+ c_{k \sigma l}\nonumber \\
 & + & \frac {4J_l\sqrt{2s}}{\sqrt{N}}\sum\limits_{k q p \in B_r } \delta (p-q-k)\cos\frac {k_x}{2} \cos\frac {k_y}{2} \cos\frac {k_z}{2} \nonumber \\ 
 & \times & \left(c_{p\downarrow l}^ + c_{q\uparrow l}a_k+c_{q\uparrow l}^ + c_{p\downarrow l}a_k^+\right) ,  \eea
with bose dispersion $\varepsilon_k$ of spin ($s=2$) iron magnons 
\be\label{FeCrS3}
\varepsilon_k  =  2sJ^A\left (3-\cos k_x-\cos k_y-\cos k_z\right),\ee \\
 and fermi $\varepsilon_{k \sigma l}$ dispersions of chromium electrons
\bea\label{FeCrS4}
\varepsilon_{k \uparrow l} & = & -2t\left ( \cos k_x+\cos k_y+\cos k_z \right)+\frac {8sJ_l-H}{2} \nonumber \\
\\
\varepsilon_{k \downarrow l} & = & -2t(\cos k_x+\cos k_y+\cos k_z)-\frac {8sJ_l-H}{2}. \nonumber
\eea    
The bosons $( a_k^+ a_k )$ are introduced by means of Holstein-Primakoff representation of the spin-2 operators of localized iron electrons.
    
To proceed we account for the spin fluctuations of iron, and in static approximation obtain three effective four-fermion theories with Hamiltonians $h^{eff}_l$, 
\bea\label{}
h^{eff}_l & = & \sum\limits_{k\in B_r \sigma} \varepsilon_{k \sigma l} c_{k \sigma l}^+ c_{k \sigma l} \\
& - & \frac 1N \sum\limits_{k_i p_i \in B_r} \delta (k_1-k_2-p_1+p_2) \nonumber \\
& \times & V_{k_1-k_2}^l c_{k_1\downarrow l}^+c_{k_2\uparrow l}c_{p_2\uparrow l}^+c_{p_1\downarrow l} \nonumber \eea
with potentials
\be\label{IFer}
V_{k}^l= \frac { J^2_l  (1+\cos k_x )(1+\cos k_y )(1+ \cos k_z )}{J^A\left ( 3-\cos k_x-\cos k_y-\cos k_z \right)} \ee
The Hamiltonians in the Hartree-Fock approximation are
\be\label{}
h_l^{HF}= \sum\limits_{k\in B_r}\left[ \varepsilon_{k \sigma l} c_{k \sigma l}^+ c_{k \sigma l}+\Delta_{k l} c_{-k\downarrow l}^+c_{k\uparrow l}+\Delta_{k l}^+c_{k\uparrow l}c_{-k\downarrow l}\right],\ee
with gap functions
\be\label{}
\Delta_{k l}=\frac 1N \sum\limits_{p\in B_r}<c_{-p\uparrow l}c_{p\downarrow l}> V_{p-k }^l \ee
In terms of  Bogoliubov excitations $\alpha_l^+,\alpha_l,\beta_l^+,\beta_l$, with dispersions
\bea\label{IFerri10}
E^{\alpha}_{k l} & = & \frac 12 \left[\varepsilon_{k\uparrow l}-\varepsilon_{k\downarrow l}+\sqrt{(\varepsilon_{k\uparrow l}+\varepsilon_{k\downarrow l})^2+4|\Delta_{k l}|^2}\right] \\
E^{\beta}_{k l} & = & \frac 12 \left[-\varepsilon_{k\uparrow l}+\varepsilon_{k\downarrow l}+\sqrt{(\varepsilon_{k\uparrow l}+\varepsilon_{k\downarrow l})^2+4|\Delta_{k l}|^2}\right]. \nonumber\eea
the gap equations have the form
\bea\label{IFerri11}\nonumber
\Delta_{k l}= & - & \frac 1N \sum\limits_{p\in B_r}V_{k+p}^l\frac {\Delta_{p l}}{\sqrt{(\varepsilon_{p\uparrow l}+\varepsilon_{p\downarrow l})^2+4|\Delta_{p l}|^2}} \\
& \times & \left(1-<\alpha^+_{p l}\alpha_{p l}>-<\beta^+_{p l}\beta_{p l}>\right), \eea
where $<\alpha^+_{p l}\alpha_{p l}>$ and $<\beta^+_{p l}\beta_{p l}>$ are fermi functions for Bogoliubov fermions.

\begin{figure}[!ht]
%\centering\includegraphics[width=4in]{(fig2)Sc-FeCrS-gap-H.pdf}
\epsfxsize=\linewidth
\epsfbox{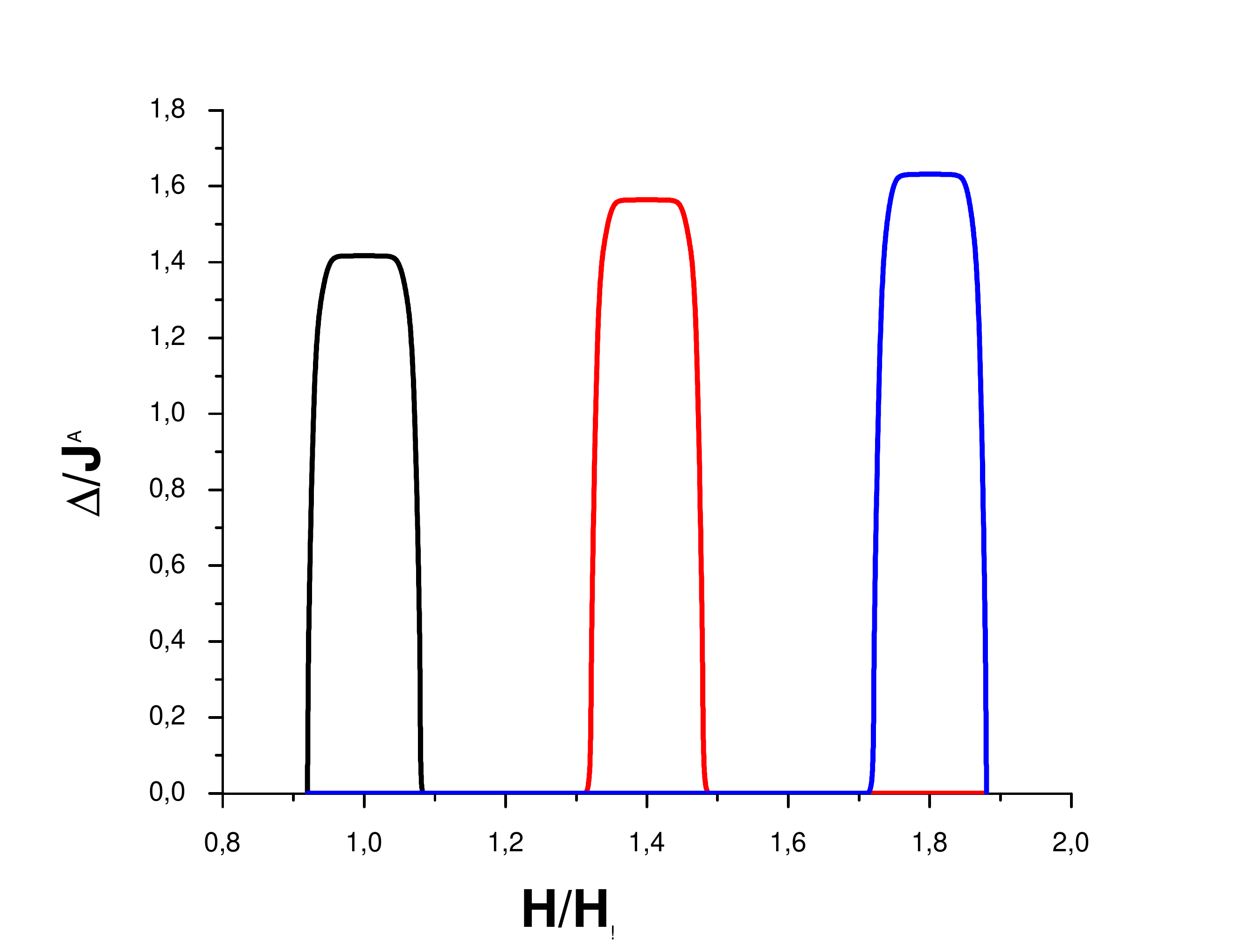}
\caption{(Color online)Sequence of superconducting states in field cooled $FeCr_2S_4$. The first state is realized near $H=H_1=2sJ_1$, the second one near $H=H_2=2sJ_2=1.4 H_1$, and the third state near $H=H_3=2sJ_3=1.8 H_1$. } \label{(fig2)}
\end{figure}

Having in mind that sublattices are simple cubic lattices and following the classifications for spin-triplet gap functions $\Delta_{-k l}=-\Delta_{k l}$, we obtained that the gap functions with $T_{1u}$ configuration \cite{RKS10}
\be\label{IFerri12} \Delta_{k l}=\Delta_l\left(\sin k_x+\sin k_y+\sin k_z) \right) \ee
are solutions of the gap equations for some values of the applied, during the preparation, magnetic field and temperature. The dimensionless gaps $\Delta_l/J^A$ at zero temperature,  as a function of $H/H_1$ where $H_1=8sJ_1$, are depicted in Fig.(\ref{(fig2)}) for parameters $J_1/J^A=2$, $J_2/J_1=1.4$, $J_3/J_1=1.8$ and $t/J^A=1$. 
%\vskip -1.cm

The applied magnetic field $H_1$ compensates the Zeeman splitting of $t_{2g}$ chromium electrons with minimum   Zeeman energy (\ref{FeCrS4}). The Fig.(\ref{(fig2)}) shows that near this value the above mentioned electrons form Cooper pairs and superconductivity onsets. Increasing the magnetic field we restore the Zeeman splitting with opposite sign and suppress the superconductivity. 
Further increasing the magnetic field, we reach $H_2=2sJ_2$ that compensates the Zeeman splitting of another $t_{2g}$ electrons. Now, the Cooper pairs are formed by the second group electrons and the superconductivity is restored near $H/H_1 = H_2/H_1 = J_2/J_1=1.4$. This process continues until the magnetic field, applied during preparation, becomes equal to Zeeman energy of the third group of chromium electrons $H_3 =2sJ_3$ and third superconductor state emerges near $H_3/H_1=1.8$. In that way we can create a sequence of superconducting states in field cooled $FeCr_2S_4$.

\begin{figure}[!ht]
%\centering\includegraphics[width=4.6in]{fig4UGe2Tx&Tsc.pdf}
\epsfxsize=\linewidth
\epsfbox{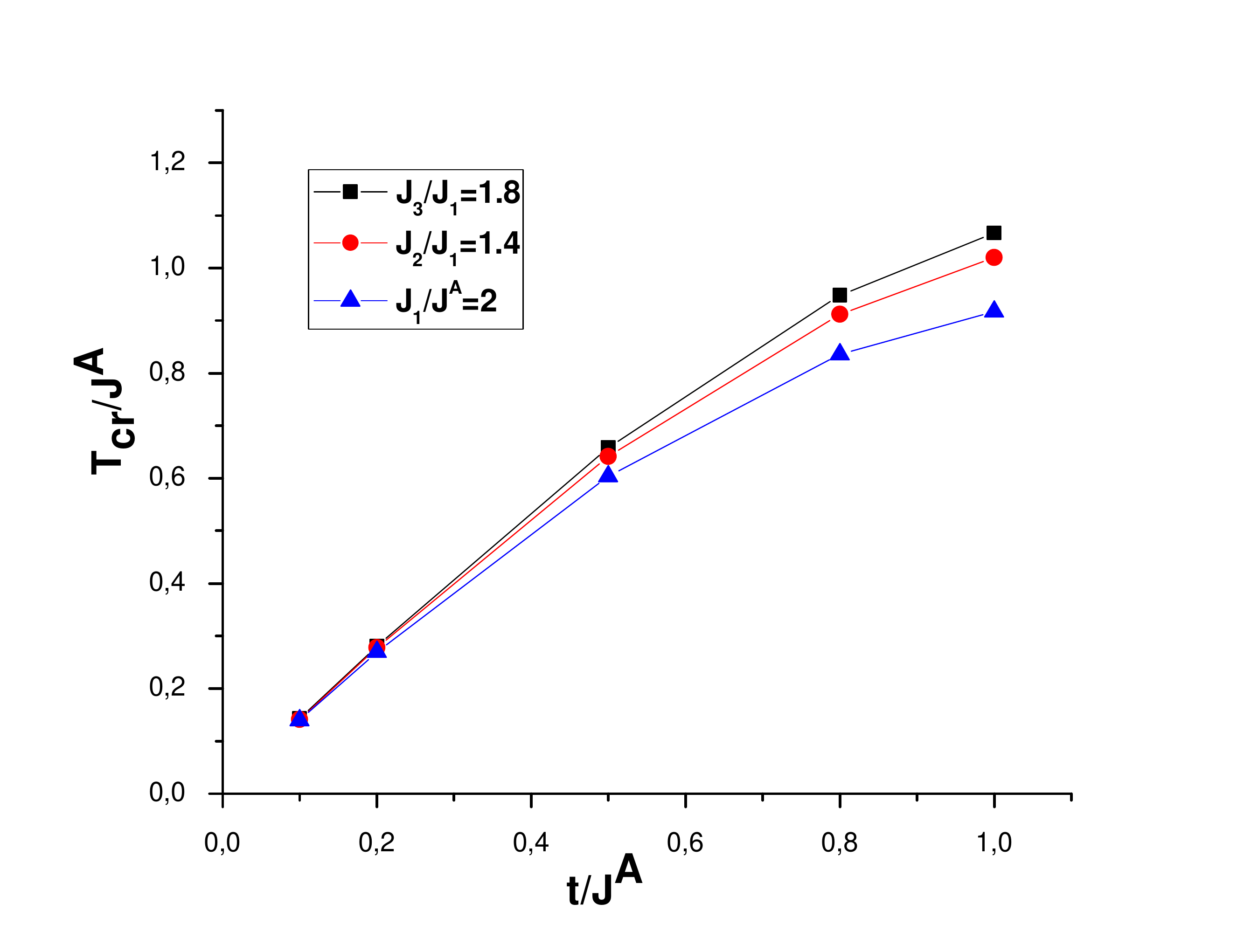}
\caption{(Color online)The dimensionless critical temperatures $T_{sc}/J^A$ as functions of dimensionless hopping parameter $t/J^A$. }\label{(fig3)}
\end{figure}

The model has many parameters. Some of them can to be extracted from the experimental data, while there is no way to define others. For example the exchange constants can be fixed from applied, during the preparation, magnetic field. Unfortunately,  the experimental articles  do not comment on this issue. We can approximately to estimate the exchange constant of iron spins $J ^ A $ by calculation the dimensionless critical temperature $T_{cr}/J^A$ for spin-2 Heisenberg model of iron magnetism. Using the renormalized spin-wave theory, one obtains 
$T_{cr}/J^A = 7.4$, therefore $J^A=23 K$. The  hopping parameter $t$ remains undefined. To shed light on critical temperature values $T_{sc}$ we calculate the dimensionless temperature $T_{sc}/J^A$ as a function of the ratio $t/J^A$. The results are depicted in Fig.(\ref{(fig3)}).
%\vskip -0.6 cm

The  Fig.(\ref{(fig3)}) shows that  increasing the ration  $t/J^A$ increases the critical temperature $T_{sc}/J^A$. We calculated that for fixed value $J^A=23 K$ and 
small value of the ratio $t/J^A=0.1$ the critical temperature is $T_{sc}=3.2 K$, while for $t/J^A=1$ we obtained $T_{sc}=23.5 K$. The figure shows that the impact of the magnetic field, applied during the preparation, is weak. The effect becomes significant when $t/J^A>>1$.

In summary, we have predicted the possibility to synthesize three different superconductors applying different magnetic field during preparation of field cooled $FeCr_2S_4$. The difficult moment is to applied a field which compensates the Zeeman splitting of one of the chromium electrons. An useful guidance to do that are the curves depicted in figure (\ref{(fig1)}).    
The curves FC1, FC2 and FC4 illustrate the three cases when one of chromium electrons is with compensated Zeeman splitting. They are well separated from the others which permits the exact choice of the applied, during preparation, magnetic field.

Partial order and partial order transition are basic notions in our study. In the present paper temperature $T^*=160 K $ just below Curie is a partial order transition temperature above which only iron ions contribute the magnetization of system, while below it one of the chromium electrons onsets magnetic moment. This is more evident differentiating the curves of ZFC and FC compounds. The separation is attributed to magnetic domain dynamics in \cite{Tsurkan10}. In the same article  small dip of the curve at  
approximately $10 K$ is interpreted as onset of long-range orbital order, while in our study, at this temperature onsets magnetization another chromium electron.

The spinel $FeCr_2S_4$ is well studied compound, but superconductivity has not been observed. The explanation is very simple. The focus of research is on colossal magnetoresistance effect (CMR) \cite{RCK1997,Yang00} and on the existence of multiferroic phase. The investigation of CMR requires measurements of resistivity as a function of temperature, but they are realized for ZFC materials only. In the rare cases, when measuring the resistivity of FC compounds, there is no clarity about the applied magnetic field. This can be achieved by simultaneous study spontaneous magnetization and resistivity as functions of temperature, which would lead to  correct choice of field. 

The third characteristic temperature of the system is $T=60 K$. It is reported \cite{JB14,Lin14} that below $50 K$ the system undergoes a transition to non-collinear ferrimagnetism. Important consequence is the emergence of multiferroic phase below $10 K$. In the present paper the system undergoes partial order transition at 60 K. Below this temperature one of the chromium electrons onsets magnetic order. This is more evident for FC compounds Fig.(\ref{(fig1)}). One expects that applying magnetic field, during preparation, the non-collinear order is suppressed and collinear ferrimagnetism is restored. Hence, the model considered in the paper is appropriate for study field-cooled $ FeCr_2S_4 $.


\vskip 0.6cm
\begin{thebibliography}{99}
%
\bibitem{UGe2000} S. Saxena, P. Agarwal, K. Ahilan, F. M. Grosche, R. Haselwimmer,
M. Steiner, E. Pugh, I. Walker, S. Julian, P. Monthoux, G. Lonzarich, A.
Huxley, I. Sheikin, D. Braithwaite, and J. Flouquet, Nature (London) {\bf406}, 587 (2000).
\bibitem{Huxley2001} A. Huxley, I. Sheikin, E. Ressouche, N. Kernavanois, D. Braithwaite, R. Calemczuk, and J. Flouquet, 
Phys. Rev. {\ B 63}, 144519 (2001).
\bibitem{Tateiwa2001b} N. Tateiwa, K. Hanazono, T. C. Kobayashi, K. Amaya,
T. Inoue, K. Kindo, Y. Koike, N. Metoki, Y. Haga, R. Settai, and Y. Onuki, J. Phys. Soc. Jpn {\bf 70}, 2876 (2001).
\bibitem{Motoyama2001} G. Motoyama, S. Nakamura, H. Kadoya, T. Nishioka, and N. K. Sato,Phys. Rev. B 65, 020510(R) (2001).
\bibitem{Huxley2002} C. Pfleiderer and A. D. Huxley, Phys. Rev. Lett., {\bf 89}, 147005 (2002).
\bibitem{Pfleiderer2009} Christian Pfleiderer, Rev. Mod. Phys. {\bf 81}, 1551 (2009).
\bibitem{Aoki2019} Dai Aoki, Kenji Ishida, and Jacques Flouquet, J. Phys. Soc. Jpn. {\bf 88}, 022001 (2019).  
\bibitem{Ashcroft68} N. W. Ashcroft, Phys. Rev. Lett. {\bf 21}, 1748 (1968).
\bibitem{Ashcroft04} N. W. Ashcroft, Phys. Rev. Lett. {\bf 92}, 187002 (2004).
\bibitem{Drozdov15} A. P. Drozdov, M. I. Eremets, I. A. Troyan, V. Ksenofontov, and S. I Shylin, Nature {\bf 525}, 73 (2015).
\bibitem{Drozdov19} A. P. Drozdov, M. I. Eremets, I. A. Troyan, V. Ksenofontov, and S. I Shylin, Nature {\bf 569}, 528 (2019).
\bibitem{Somayazulu19} Maddury Somayazulu, Muhtar Ahart, Ajay K. Mishra, Zachary M. Geballe, Maria Baldini,
Yue Meng, Viktor V. Struzhkin, and Russell J. Hemley1,  Phys. Rev. Lett. {\bf 122}, 027001 (2019).
\bibitem{Snider20} Elliot Snider, Nathan Dasenbrock-Gammon, Raymond McBride, Mathew Debessai, Hiranya Vindana, Kevin Vencatasamy, Keith V. Lawler, Ashkan Salamat, and Ranga P. Dias, Nature {\bf 586}, 373 (2015).
\bibitem{Bednorz86} J. G. Bednorz and K. A. Mueller,  Z. Phys. B: Condens. Matter {\bf 64}, 189 (1986).
\bibitem{Tokura89} Y. Tokura et al, Nature (London) {\bf 337} 345 (1989).
\bibitem{Ginsberg89} D. M. Ginsberg (Eds.) Physical Properties of High Temperature Superconductivity,{\bf  Vols. 1–5}, World Scientific, Singapore, (1989 - 1996).
\bibitem{Dagotto94} E. Dagotto, Rev. Mod. Phys.{\bf 66} 763 (1994).  
\bibitem{Anderson97} P.W. Anderson,  \textit{The Theory of Superconductivity in the High-$T_c$ Cuprates}, Princeton University Press, Princeton, (1997).
\bibitem{Lee06} Patrick Lee and Naoto Nagaosa, Rev. Mod. Phys., {\bf 78} 17 (2006).
\bibitem{Fischer75} O. Fischer, M. Decroux, S. Roth and M. Sergent, J. Phys. {\bf C 8}, L474 (1975).
\bibitem{Fischer84} H. W. Meul, C. Rossel, M. Decroux, and 0. Fischer, G. Remenyi and A. Briggs, Phys. Rev. Lett., {\bf 54}, 2541 (1985).
\bibitem{JP62} V. Jaccarino and M. Peter, Phys. Rev. Lett. {\bf 9}, 290 (1962).
\bibitem{Lin85} C. L. Lin, J. Teter, J. E. Crow, T. Mihalisin, J. Brooks, A. I. Abou-Aly, and G. R. Stewart,  Phys. Rev. Lett., {\bf 53}, 497 (1984).
\bibitem{Shimahara02} H. Shimahara, J. Phys. Soc. Jpn. {\bf 71}, 713 (2002).
\bibitem{Hardy05} F. Hardy and A. D. Huxley, Phys. Rev. Lett. {\bf 94}, 24700  (2005).
\bibitem{Sheikin05} F. L\'evy, I. Sheikin, B. Grenier, and A. Huxley, Science {\bf 309},  1343 (2005).
\bibitem{Uji01} S. Uji, H. Shinagawa, T. Terashima, T. Yakabe, Y. Terai, M. Tokumoto, A. Kobayashi, H. Tanaka, and H. Kobayashi, Nature (London) {\bf 410}, 908 (2001).
\bibitem{Balicas01} L. Balicas, J. S. Brooks, K. Storr, S. Uji, M. Tokumoto, H. Tanaka, H. Kobayashi, A. Kobayashi, V. Barzykin, and L. P. Gor’kov, Phys. Rev. Lett.{\bf 87}, 067002 (2001).
\bibitem{Konoike04} T. Konoike, S. Uji, T. Terashima, M. Nishimura, S. Yasuzuka, K. Enomoto, H. Fujiwara, B. Zhang, and H. Kobayashi, Phys. Rev. {\bf B 70}, 094514 (2004).
\bibitem{Diep87} P. Azaria, H. T. Diep, and H. Giacomini, Phys. Rev. Lett. {\bf 59}, 1629 (1987).
\bibitem{Larkin66} V. G. Vaks, A. I. Larkin, and Y. N. Ovchinnikov, JETP Letters. {\bf 22}, 820 (1966).
\bibitem{Diep04} H. T. Diep, Ed., \textit{Frustrated Spin Systems}, World Scientific (2004)
\bibitem{Diep97} R. Quartu and H. T. Diep, Phys. Rev. {\bf B 55}, 2975 (1997).
\bibitem{POexp04} J. R. Stewart, G. Ehlers, A. S. Wills, S. T. Bramwell, and J. S. Gardner, J. Phys.: Condens. Matter {\bf 16}, L321 (2004).
\bibitem{Karchev08} N. Karchev, Phys. Rev. {\bf B 77}, 012405 (2008). 
\bibitem{spinel+} K. Adachi, T. Suzuki, K. Kato, K. Osaka, M. Takata and T. Katsufuji, Phys. Rev. Lett. {\bf 95}, 197202 (2005).
\bibitem{spinelFeCr2S4} Zhaorong Yang, Shun Tan, Zhiwen Chen, and Yuheng Zhang,  Phys. Rev. {\bf B 62}, 13872 (2000).
\bibitem{spinelCv1} H. D. Zhou, J. Lu, and C. R. Wiebe, Phys. Rev. {\bf B 76}, 174403 (2007).
\bibitem{spinel08} V. O. Garlea, R. Jin, D. Mandrus, B. Roessli, Q. Huang, M. Miller, A. J. Schultz, and S. E. Nagler, Phys. Rev. Lett. {\bf 100}, 066404 (2008).
\bibitem{spinel++} S-H. Baek, K-Y. Choi, A. P. Reyes, P. L. Kuhns, N. J. Curro, V. Ramanchandran, N. S. Dalal, H. D. Zhou, and C. R. Wiebe,
J. Phys.: Condens. Matter {\bf 20}, 135218 (2008).
\bibitem{spinel11b} Kim Myung-Whun, J. S. Kim, T. Katsufuji, and R. K. Kremer, Phys. Rev. {\bf B 83}, 024403 (20011).
\bibitem{spinel11a} A. Kiswandhi, J. S. Brooks, J. Lu, J. Whalen, T. Siegrist, and H. D. Zhou, Phys. Rev. {\bf B 84}, 205138 (2011).
\bibitem{spinel11c} A. Kismarahardja, J. S. Brooks, A. Kiswandhi, K. Matsubayashi, R. Yamanaka, Y. Uwatoko, J. Whalen,
T. Siegrist, and H. D. Zhou, Phys. Rev. Lett. {\bf 106}, 056602 (2011).
\bibitem{spinel12a}Q. Zhang, K. Singh, F. Guillou, C. Simon, Y. Breard, V. Caignaert, and V. Hardy, Phys. Rev. {\bf B 85}, 054405 (2012).
\bibitem{spinel12b} Y. Nii, H. Sagayama, T. Arima, S. Aoyagi, R. Sakai, S. Maki, E. Nishibori, H. Sawa, K. Sugimoto,
H. Ohsumi, and M. Takata, Phys. Rev. {\bf B 86}, 125142 (2012).
\bibitem{spinel+1} Z. H. Huang, X. Luo, S. Lin, Y. N. Huang, L. Hu, L. Zhang, Y. P.Sun, Solid State Communications {\bf 159}, 88 (2013).
\bibitem{spinel+2} Z. H. Huang, X. Luo, L. Hu, S. G. Tan, Y. Liu, B. Yuan, J. Chen, W. H. Song, and Y. P. Sun, Journal of Applied Physics {\bf 115}, 034903 (2014).
\bibitem{spinel+3} Dina Tobia, Juli$\acute{a}$n Milano, Maria Teresa Causa and Elin L. Winkler,
J. Phys.: Condens. Matter {\bf 27}, 016003 (2015).
\bibitem{Yang00} Zhaorong Yang, Shun Tan, Zhiwen Chen, and Yuheng Zhang, Phys. Rev. {\bf B 62}, 13872 (2000).
\bibitem{Tsurkan10} V. Tsurkan, O. Zaharko, F. Schrettle, Ch. Kant, J. Deisenhofer, H.-A. Krug von Nidda, V. Felea, P. Lemmens,
J. R. Groza, D. V. Quach, F. Gozzo and A. Loidl, Phys. Rev. {\bf B 81}, 184426 (2010).
\bibitem{Yto11} Masakazu Ito, YujiNagi, NaotoshiKado, ShinpeiUrakawa, TakuroOgawa, AkihiroKondo,
Keiichi Koyama, KazuoWatanabe, KoichiKindo, J. Mag. Mag. Mater, {\bf 323}, 3290 (2011).
\bibitem{Lin14} L. Lin, H. X. Zhu, X. M. Jiang, K. F. Wang, S. Dong, Z. B. Yan, Z. R. Yang, J. G. Wan, and J.-M. Liu, Sci. Rep., {\bf4}, 6530 (2014).
\bibitem{RKS10} S. Raghu, S. A. Kivelson, \and D. J. Scalapino, Phys. Rev. {\ bf B 81}, 224505 (2010).
\bibitem{RCK1997} A. P. Ramirez, R. J. Cava, and J. Krajewsky, Nature {\bf 386}, 156 (1997).
\bibitem{JB14} J. Bertinshaw, C. Ulrich, A.G\"unther, F. Schrettle, M. Wohlauer, S. Krohns, M. Reehuis, A. J. Studer, M. Avdeev, D. V. Quach, J. R. Groza, V. Tsurkan, A. Loidl, and J. Deisenhofer, Sci. Rep., {\bf 4}, 6079 (2014).






 \end{thebibliography}
\end{document}